\documentclass[aip,rsi,amsmath,amssymb,reprint]{revtex4-1}
\usepackage{graphicx}
\usepackage{hyperref} 
\usepackage{amssymb}
\usepackage{color}
\usepackage{filecontents}

\DeclareUnicodeCharacter{2212}{\textendash}
\begin{document}

\preprint{AIP/123-QED}

\title{Re-defining the concept of hydration water in water under soft confinement}

\author{Fausto Martelli}
\affiliation{IBM Research Europe, Hartree Centre, Daresbury, WA4 4AD, United Kingdom}
\email{fausto.martelli@ibm.com}
\affiliation{Department of Physics and CNR Institute of Complex Systems, Sapienza University of Rome, P.le Aldo Moro 2, 00185 Roma, Italy}
\author{Carles Calero}
\affiliation{Secci\'o de
F\'isica Estad\'istica i Interdisciplin\`aria--Departament de F\'{i}sica de la Mat\`{e}ria Condensada, Universitat de Barcelona, C. Mart\'{i} i Franqu\`{e}s 1, 08028 Barcelona, Spain}
\email{carles.calero@ub.edu; gfranzese@ub.edu}
\affiliation{Institut de Nanoci\`{e}ncia i Nanotecnologia (IN2UB), Universitat de Barcelona, C. Mart\'{i} i Franqu\`{e}s 1, 08028 Barcelona, Spain}
\author{Giancarlo Franzese}
\affiliation{Secci\'o de
F\'isica Estad\'istica i Interdisciplin\`aria--Departament de F\'{i}sica de la Mat\`{e}ria Condensada, Universitat de Barcelona, C. Mart\'{i} i Franqu\`{e}s 1, 08028 Barcelona, Spain}
%\email{carles.calero@ub.edu; gfranzese@ub.edu}
\affiliation{Institut de Nanoci\`{e}ncia i Nanotecnologia (IN2UB), Universitat de Barcelona, C. Mart\'{i} i Franqu\`{e}s 1, 08028 Barcelona, Spain}

\keywords{keywords here}

\begin{abstract}
 Water shapes and defines the properties of biological systems. Therefore, understanding the nature of the mutual interaction between water and biological systems is of primary importance for a proper assessment of biological activity and the development of new drugs and vaccines. A handy way to characterize the interactions between biological systems and water is to analyze their impact on water density and dynamics in the proximity of the interfaces. It is well established that water bulk density and dynamical properties are recovered at distances in the order of $\sim1$~nm from the surface of biological systems. Such evidence led to the definition of \emph{hydration} water as the thin layer of water covering the surface of biological systems and affecting-defining their properties and functionality. Here, we review some of our latest contributions showing that phospholipid membranes affect the structural properties and the hydrogen bond network of water at greater distances than the commonly evoked $\sim1$~nm from the membrane surface. Our results imply that the concept of hydration water should be revised or extended, and pave the way to a deeper understanding of the mutual interactions between water and biological systems.
\end{abstract}


\maketitle

\section{Introduction}
Water is a peculiar substance characterized by a plethora of dynamic and thermodynamic anomalies that make it the only liquid capable to sustain life as we know it \cite{WaterandLife, Chaplin, Ball:2008aa}. For example, the very large heat capacity allows water to absorb and release heat at much slower rates compared to similar materials like silica. As a consequence, water acts as a thermostat that regulates the temperature of our bodies and, overall, of our planet sheltering us from otherwise lethal daily and seasonal temperature variations. Water has also a very low compressibility, that allows blood to be pumped without crystallizing down to the most peripherals and tight vessels delivering oxygen. 
Nonetheless, water stabilizes proteins and DNA restricting the access to  unfolded states, and shapes the basic structure of cells membranes. Cells membranes are very complex systems made of a large number of components, including proteins, cholesterol, glycolipids and ionic channels among others, but their framework is provided by phospholipid molecules forming a bilayer. Being solvated by water, the hydrophilic heads of the phospholipid molecules are exposed to the surrounding solvent molecules, while the hydrophobic tails are arranged side by side hiding from water and extending in the region between two layers of heads. Stacked membranes are important constituents in several biological structures, including endoplasmic reticulum and Golgi apparatus, that processes proteins for their use in animal cells, or thylakoid compartments in chloroplasts and cyanobacteria, involved in photosynthesis. When in contact with membranes, water modulates their fluidity and mediates the interaction between different membranes as well as between membranes and solutes (ions, proteins, DNA, etc.), regulating cell-membrane tasks such as, e.g., transport and signaling functions \cite{hamley}. A thin layer of water, with a thickness of only $\sim1$~nm corresponding to a couple of molecular diameters, hydrates biological systems and is therefore called \emph{biological}, or \emph{hydration} water \cite{Zhong:2011ab}. So far, it has been thought that hydration water is directly responsible for the proper functioning of biological systems \cite{Chaplin}, although many issues are still open \cite{Zhong:2011ab}.

Several experimental techniques have been adopted to study the interaction between hydration water molecules and membrane surfaces. Insights on the orientation of water molecules and on their order have been obtained from vibrational sum frequency generation spectroscopy and nuclear magnetic resonance (NMR) experiments~\cite{konig_1994,chen_2010}. Evidences of enhanced hydrogen bonds (HBs) established between water molecules and the phospholipid heads have been described in experimental investigations from infrared spectroscopy~\cite{binder_2003,chen_2010}. Nonetheless, far-infrared spectroscopy has shown that resonance mechanisms entangle the motion of phospholipid bilayers with their hydration water~\cite{dangelo_2017}. Such complex interactions between water molecules and hydrophobic heads cause perturbations in the dynamical properties of water. NMR spectroscopy has reported a breakdown of the isotropy on the lateral and normal diffusion of water molecules with respect to the surface~\cite{Volke1994,Wassall_BiophysJ1996}, and rotational dynamics has been the focus of several experimental investigations using ultrafast vibrational spectroscopy~\cite{Zhao_Fayer_JACS2008}, terahertz spectroscopy~\cite{Tielrooij_BiophysJ2009} and neutron scattering~\cite{Trapp_JCP2010}.

Atomistic molecular dynamics (MD) simulations have also been widely adopted to inspect the microscopic details of hydration water (with the obvious drawback of relying on a particular simulation model). The dynamical slow-down of water dynamics due to the interaction with phospholipid membranes reported in NMR experiments~\cite{Volke1994,Wassall_BiophysJ1996} has been confirmed in MD simulations~\cite{Berkowitz_chemrev2006,Bhide_JCP2005}. MD simulations have also provided important insights on the molecular ordering and rotation dynamics in water solvating phospholipid headgroups~\cite{Berkowitz_chemrev2006,pastor_1994}, as well as in quantifying --introducing correlation functions-- the decay of water orientational degrees of freedom~\cite{Zhang_Berkowitz_JPhysChemB2009,Gruenbaum_JChemPhys_2011,calero_2016,martelli_fop,2018arXiv181101911S}. 

\begin{figure}%[ht]
  \begin{center}
   \includegraphics[scale=.25]{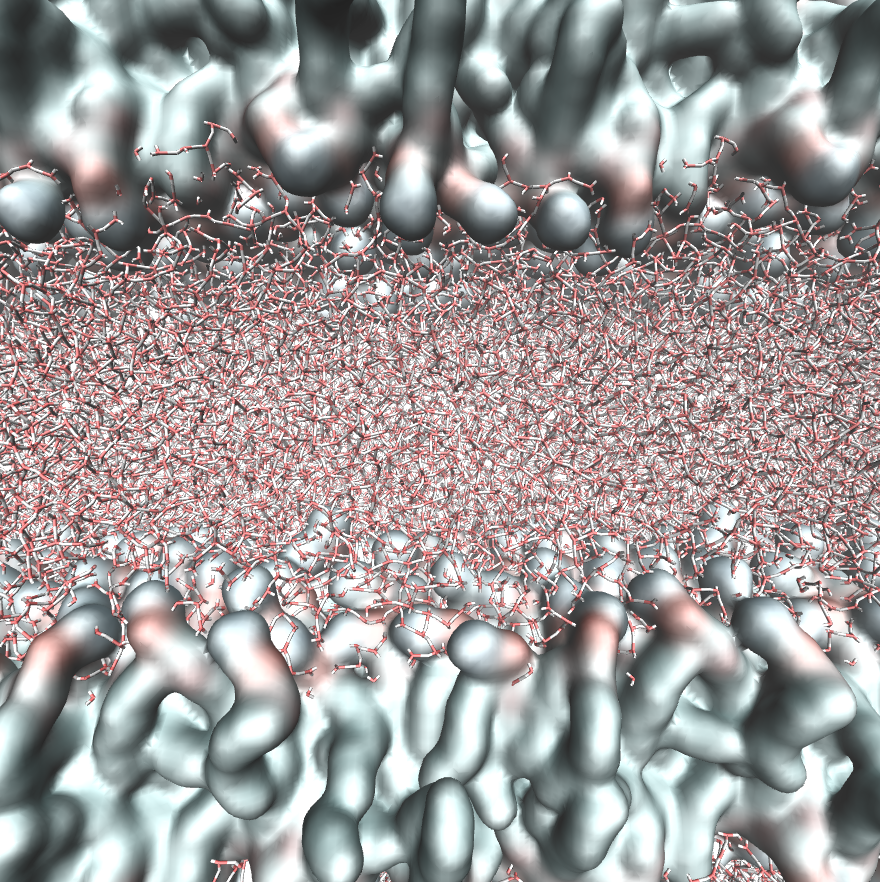}
    \caption{\label{fig:fig1} 
    Representative snapshot of a molecular system composed by water molecules (sticks) and DMPC leaflets (blur fields).}
  \end{center}
\end{figure}

We here review some of our recent computational investigations on water nanoconfined between stacked phospholipid membranes, reporting evidences that the membrane affects the structural properties of water and its hydrogen bond network at distances much larger than the often invoked $\sim1$~nm.
Our results are the outcome of MD simulations of water nanoconfined in phospholipid membranes. Water is described via a modified TIP3P~\cite{tip3p_1} model of water. As a typical model membrane, we have used 1,2-Dimyristoyl-sn-glycero-3-phosphocholine (DMPC) lipids. The DMPC is a phospholipid with a hydrophobic tail  formed of two myristoyl chains and a 
hydrophilic head, containing a phosphate and a choline,
where the N atom interacts mostly with water oxygen atoms and the P atom interacts mostly with the hydgrogen atoms.
Choline-based phospholipids are ubiquitous in cell membranes and commonly used in drug-targeting liposomes~\cite{hamley}. 
In Fig.~\ref{fig:fig1} we report a representative snapshot of the water-DMPC system. 
%Water molecules are shown as sticks, while the DMPC molecules are represented as blur fields.

\begin{figure}%[ht]
  \begin{center}
   \includegraphics[scale=.6]{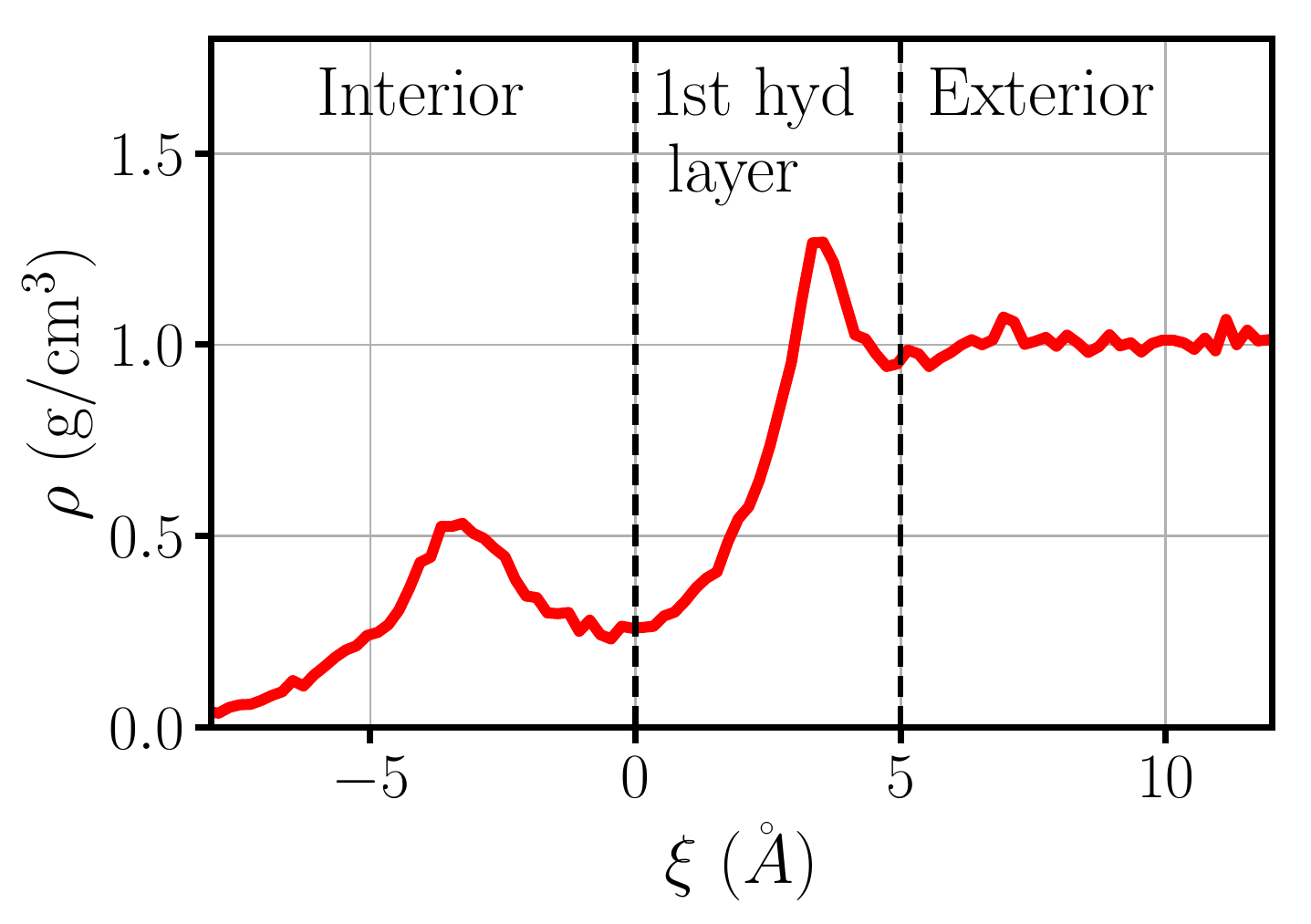}
    \caption{\label{Density_half} 
Density profile $\rho$ of water molecules as a function of the instantaneous local distance $\xi$ from the membrane interface at ambient conditions ($T=303$ K, average pressure 1 atm, corresponding to bulk density $\rho=1$g/cm$^3$) and with at hydration level,  defined as the number of water molecules per phospholipid, $\omega=34$. Water at $\xi<0$ belongs to the interior of the membrane, while that at $\xi>5$\AA\ has the same density as the bulk and can be associated to the exterior of the membrane. The density of water at $0<\xi<5$\AA\ shows a clear maximum revealing the presence of a hydration layer \cite{calero_2016}. At higher density we observe more than one hydration layer.}
  \end{center}
\end{figure}

As observed in Ref.\cite{martelli_fop}, at ambient conditions the density profile of water molecules as function of the distance with respect to the average position of the phosphorus atoms in the DMPC lipids displays no layered structure. In fact, due to the thermal fluctuations, it forms a smeared out interface that is $\sim 1$ nm wide, based on the phospholipid head density \cite{martelli_fop}. However, the interface forms instantaneous layers that can be revealed if, following Pandit et al.
\cite{pandit_algorithm}, we consider the instantaneous local distance $\xi$, defined as the distance of each water molecule from the closest cell of a Voronoi tessellation centered on the phosphorous and nitrogen atoms of the phospholipid heads (Fig.~\ref{Density_half})~\cite{calero_2016}.

\section{Dynamics}
\begin{figure}%[ht]
  \begin{center}
   \includegraphics[scale=.55]{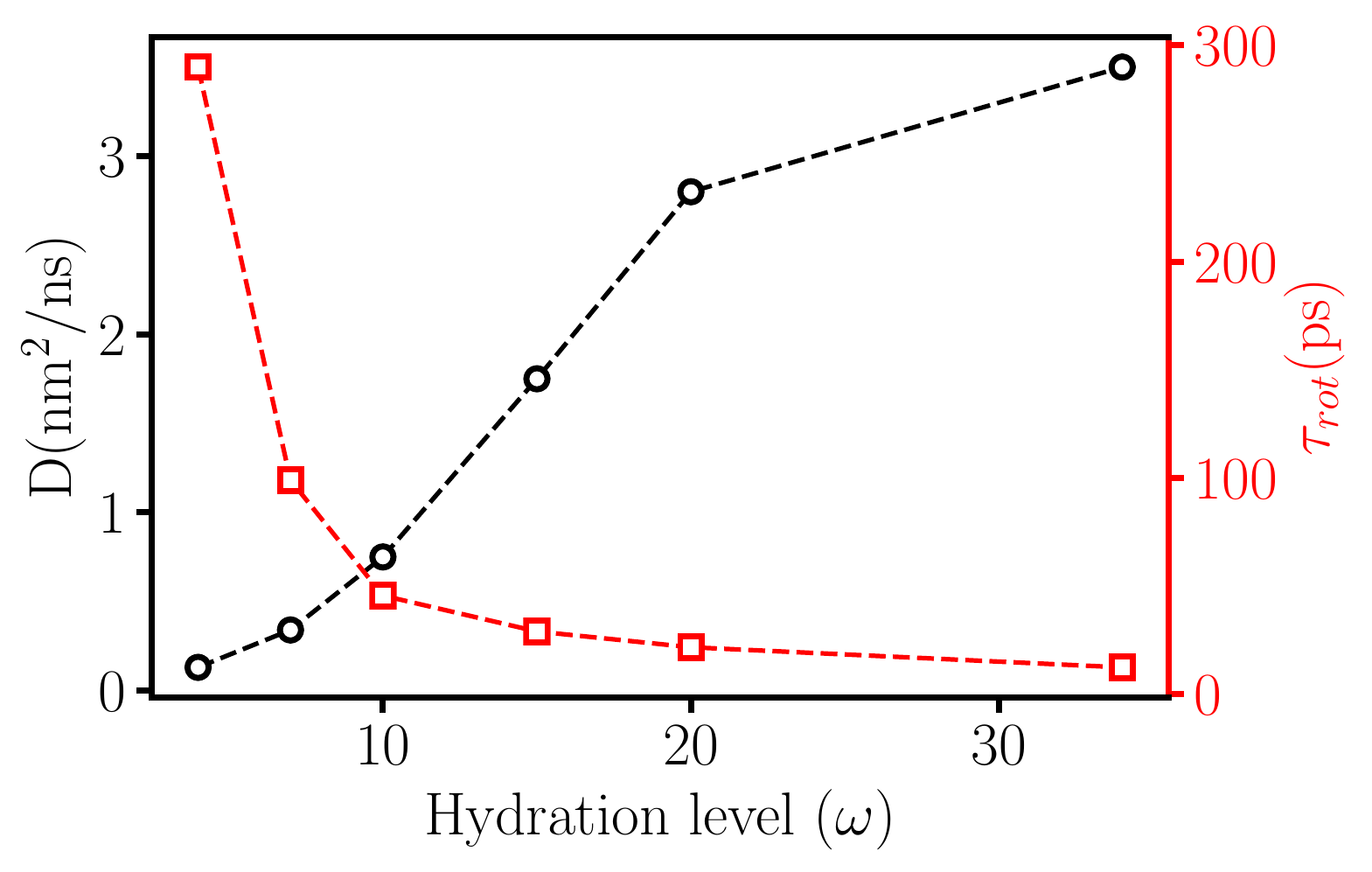}
    \caption{\label{D_Tau_hidration} 
    Dynamics of  water molecules between stacked phospholipid bilayers at different hydration level $\omega$ at ambient conditions: Diffusion coefficient $D_\parallel$ of water molecules projected on the plane of the membrane (black circles, left vertical axis); Rotational relaxation time $\tau_{rot}$ of all the water in the system (red squares, right vertical axis). Lines are guides for the eyes.}
  \end{center}
\end{figure}

Numerical simulations have shown that hydration water suffers a dramatic slow down not just in stacked phospholipids~\cite{Rog_ChemPhysLett2002,Lopez_JPhysChemB2004,Berkowitz_chemrev2006,Bhide_JCP2005,Zhang_Berkowitz_JPhysChemB2009,Gruenbaum_JChemPhys_2011,pandit_algorithm,Yang_JCP2014,calero_2016,martelli_fop,calero_membranes_2019}, but also in proteins and sugars~\cite{camisasca_2018,iorio_2019,iorio_2019_2,iorio_2019_3,iorio_2020}.
Insights on the dynamical slow down can be obtained by inspecting the translational diffusion ($D_\parallel$) and rotational dynamics of hydration water molecules. The diffusion coefficient parallel to the surface of the membrane can be obtained from the linear regime reached by the mean squared displacement at sufficiently long times from the Einstein relation:
\begin{equation}
    D_\parallel\equiv\lim_{t\rightarrow\infty}\frac{\left<\left | \mathbf{r}_\parallel(t)-\mathbf{r}_\parallel(0) \right |^2 \right>}{4t}
\end{equation}
where $\mathbf{r}_\parallel(t)$ is the projection of the center of mass of a water molecule on the plane of the membrane and the angular brackets $\left<...\right>$ indicate average over all water molecules and time origins.
Using the DMPC as a model phospholipid membrane, Calero et al.~\cite{calero_2016} have found that water molecules are slowed down by an order of magnitude when the hydration level $\omega$ is reduced from 34 to 4 (Fig.\ref{D_Tau_hidration}).
This result is in qualitative agreement with experimental and other computational studies~\cite{Wassall_BiophysJ1996,Zhao_Fayer_JACS2008,Tielrooij_BiophysJ2009,Zhang_Berkowitz_JPhysChemB2009,Gruenbaum_JChemPhys_2011}. In particular, in conditions of very low hydration, the parallel diffusion is as low as $0.13$ nm$^2$/ns because water molecules interact with both the upper and the lower leaflet, hence remaining trapped. Increasing the level of hydration $\omega$, Calero et. al~\cite{calero_2016} have shown that $D_\parallel$ increases monotonically. This observation suggests that, increasing the physical separation between the leaflets, the
hydration water  acts as a screen for the electrostatic interactions between water and the leaflets. 

The decreasing interaction of hydration water with the two leaflets can also be observed inspecting the rotational dynamics of water molecules via the rotational dipolar correlation function:
\begin{equation}
    C_{\hat{\mu}}(t)\equiv\left<\hat{\mu}(t)\cdot\hat{\mu}(0)\right>
    \label{Ct}
\end{equation}
where $\hat{\mu}(t)$ is the direction of the water dipole vector at time $t$ and $\left<...\right>$ denotes the ensemble average over all water molecules and time origins. Such quantity is related to terahertz dielectric relaxation measurements used to probe the reorientation dynamics of water~\cite{Tielrooij_BiophysJ2009}. From Eq.~\ref{Ct} it is possible to define the relaxation time
\begin{equation}
    \tau_{rot}\equiv\int_0^{\infty}C_{\hat{\mu}}(t)dt
    \label{tau}
\end{equation}
which is independent on the analytical form of the correlation function $C_{\hat{\mu}}(t)$. As for $D_{\parallel}$, the rotational dynamics speeds up with the degree of hydration (Fig.\ref{D_Tau_hidration}), confirming that the interactions between hydration water and the two leaflets modify the overall water dynamics~\cite{calero_2016,Zhao_Fayer_JACS2008,Tielrooij_BiophysJ2009,Zhang_Berkowitz_JPhysChemB2009,Gruenbaum_JChemPhys_2011}. 

\begin{figure}%[ht]
  \begin{center}
   \includegraphics[scale=.6]{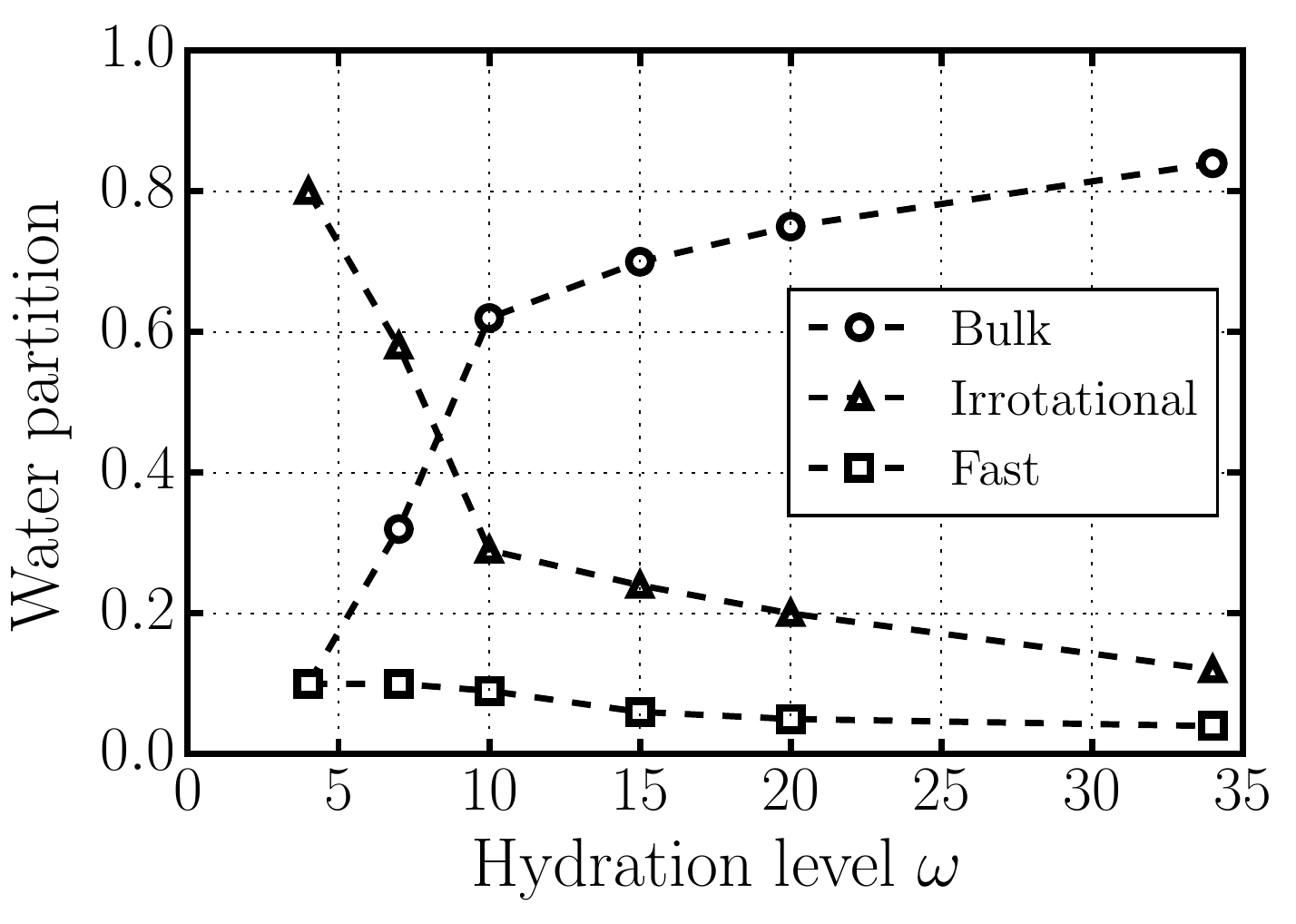}
    \caption{\label{partitioning2} 
Partition of membrane hydration water into {\em fast} (squares), irrotational (triangles) and bulk-like (circles) water molecules, following the assumption in Ref.  \cite{Tielrooij_BiophysJ2009}, as a function of the hydration level $\omega$. As discussed in Ref. \cite{calero_2016}, the assumption of the existence of  {\em fast}  water leads to inconsistencies.
}
  \end{center}
\end{figure}

To account for the rapidly relaxing signals associated with the reorientation of water molecules in experiment \cite{Righini_PRL2007}, Tielrooij et al. \cite{Tielrooij_BiophysJ2009} assumed the existence of three water species near a membrane: (i) bulk-like, with characteristic rotational correlation times of a few picoseconds; (ii) {\em fast}, with rotational correlation times of a fraction of picosecond; and (iii) irrotational, with characteristic times much larger than 10 ps. Calero et al. \cite{calero_2016} show that it is possible to analyze their simulations using this assumption (Fig. \ref{partitioning2}), however,  the resulting fitting parameters for the correlation times are  not showing any regular behavior as a function of $\omega$, questioning the existence of {\em fast} water near a membrane.  This possibility, on the other hand, cannot be ruled out completely, as it could be related to the presence of heterogeneities, such as those associated with water molecules with a single hydrogen bond to a lipid at low hydration \cite{Righini_PRL2007}.

Nevertheless, Calero et al. \cite{calero_2016} have shown that a consistent explanation of the changes in the dynamics as a function of $\omega$ is reached by observing that, upon increasing the hydration level, water  first fills completely the interior of the membrane and next accumulate in layers in the exterior region. The authors rationalized this observation observing that the inner-membrane (or interior) water has an extremely slow dynamics as a consequence of the robustness of water-lipid HBs. Moreover, the water-water HBs within the first hydration layer of the membrane slow down, with respect to bulk water, due to the reduction of hydrogen bond-switching at low hydration. As shown by Samatas et al.~\cite{2018arXiv181101911S}, these effects are emphasized when the temperature decreases: water near the membrane has a glassy-like behavior when $T = 288.6$ K, with the rotational correlation time of vicinal water, within 3 \AA\  from the membrane, comparable to that of bulk water  $\approx 30$ K colder, but with a much smaller stretched exponent, suggesting a larger heterogeneity of relaxation modes.

\begin{figure}%[ht]
  \begin{center}
   \includegraphics[scale=.3]{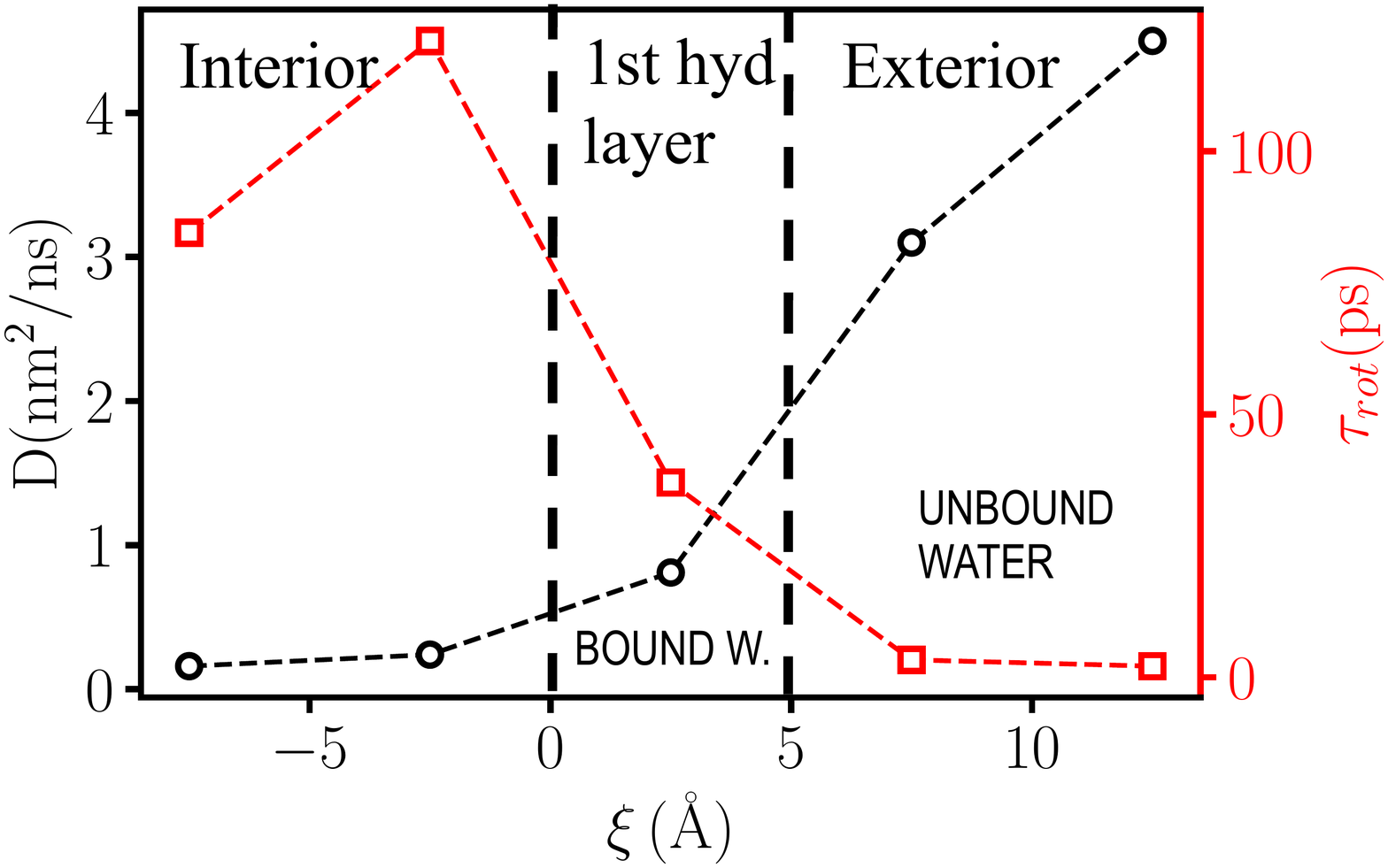}
    \caption{\label{Dcoef_Trot} 
Dynamics of  water molecules between stacked phospholipid bilayers as a function of the instantaneous local distance $\xi$ from the membrane interface at ambient conditions  and  hydration level $\omega=34$: Diffusion coefficient $D_\parallel$ of water molecules projected on the plane of the membrane (black circles, left vertical axis); Rotational relaxation time $\tau_{rot}$ of all the water in the system (red squares, right vertical axis). Lines are guides for the eyes. Vertical dashed lines at $\xi=0$ and 5 \AA\ mark the interfaces between the water within the interior of the membrane, the first hydration layer of water, and the water exterior to the membrane. The interface at $\xi=5$ \AA\ separates bound water and unbound water.}
  \end{center}
\end{figure}

Both the translational and rotational dynamics of water molecules are strongly determined by their local distance to the membrane. Calero and Franzese have recently shown \cite{calero_membranes_2019} that the hydration water within the interior of the membrane is almost immobile, the first hydration layer, with $\xi\leq  5$ \AA, is \emph{bound} to the membrane, 
and the exterior water is \emph{unbound} (Fig. \ref{Dcoef_Trot}). 
The authors have identified the existence of an interface between the bound and the unbound hydration water at which the dynamics undergoes an abrupt change:
bound water rotates 63\% less than bulk and diffuses 85\% less than bulk, while
unbound water only 20\% and 17\%, respectively.

\begin{figure}%[ht]
  \begin{center}
   \includegraphics[scale=.6]{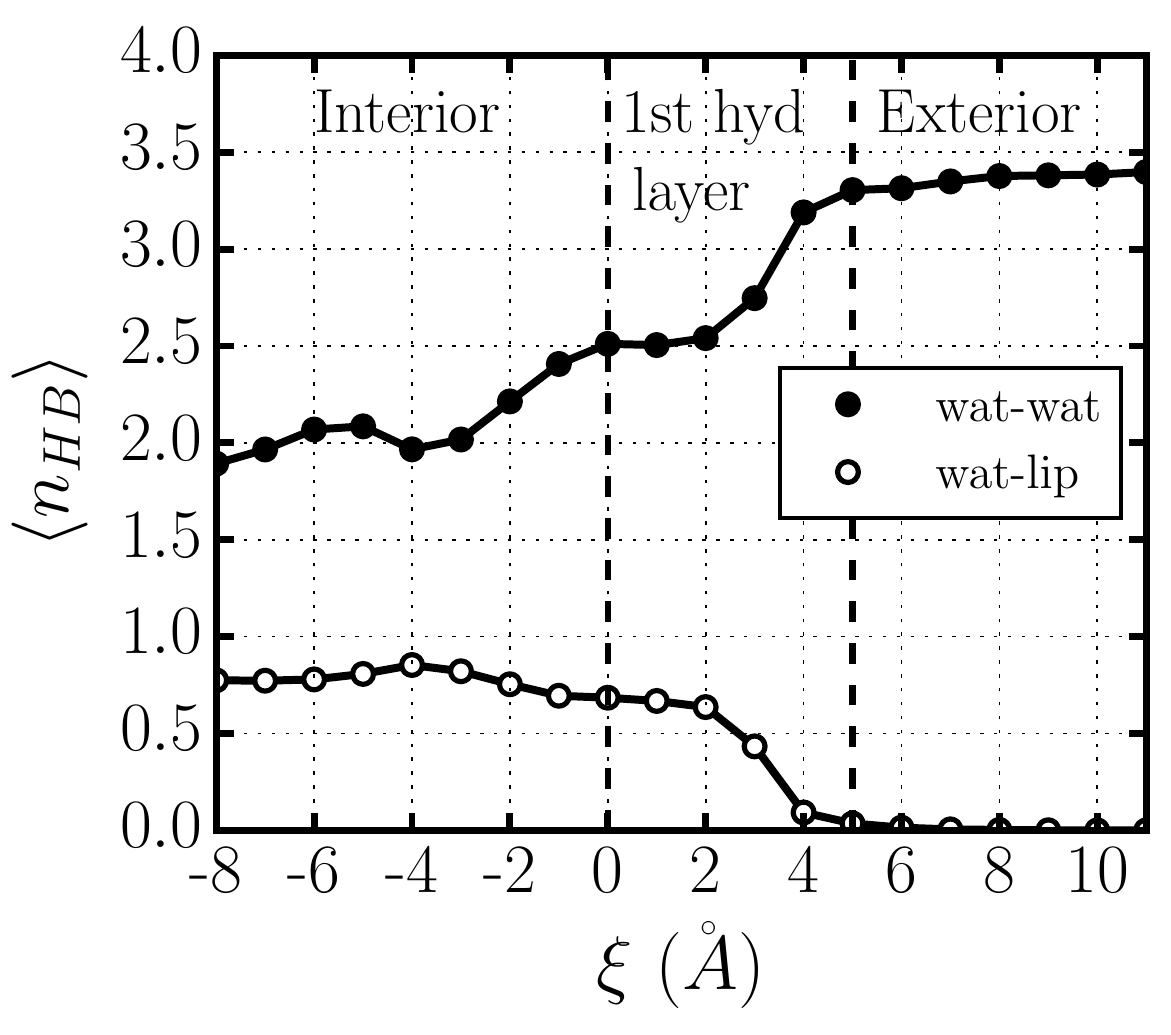}
    \caption{\label{HB_profile2} 
Average number of HBs $\langle n_{\rm HB}\rangle$ as a function of the instantaneous local distance $\xi$ from the membrane interface at ambient conditions  and  hydration level $\omega=34$. Full circles represent the HBs formed between water molecules, and empty circles the HBs formed by water molecules with selected groups of the phospholipid. Vertical dashed lines at $\xi=0$ and 5 \AA\ mark the interfaces between the interior, the first hydration layer, and the exterior water of the membrane.}
  \end{center}
\end{figure}

To rationalize the origin of the three dynamically different populations of water, (i) immobile within the membrane interior, (ii) bound in the first hydration layer, and (iii) unbound at the exterior of the membrane, Calero and Franzese have turned their attention to the investigation of the hydrogen bonds (HBs, Fig. \ref{HB_profile2}). Based on the calculation of the average number of HBs $\langle n_{\rm HB}\rangle$, they have found that the inner water is an essential component of the membrane that plays a structural role with HBs bridging between lipids, consistent with previous results \cite{Pasenkiewicz-Gierula:1997aa, lopez_2004}. In particular, Calero and Franzese have found that, in the case of a fully hydrated membrane, $\approx 45\%$ of the water- lipids HBs in the interior of the membrane are bridging between two lipids. The fraction of bridging HBs, with respect to the total number of water-lipids HBs, reduces to  approximately 1/4 within  the first hydration shell. Hence, also the  bound  water has a possible structural function for the membrane and, in this sense, can be considered as another \emph{constituent} of the membrane that regulates its properties and contributes to its stability.
Moreover, they found that unbound hydration water has no water-lipids HBs. However, even at  hydration level as low as $\omega= 4$, they find that $\approx 25\%$ of  inner water, and $\approx 18\%$ in the first hydration shell, is unbound, i.e. has only water-water HBs. This could be the possible reason why it has been hypothesized the existence of \emph{fast} water in weakly hydrated phospholipid bilayers in previous works \cite{Tielrooij_BiophysJ2009}. Nevertheless, as already discussed, Calero and Franzese clearly showed that unbound water is definitely not fast, being at least one order of magnitude slower than bulk water. 

\begin{figure}%[ht]
  \begin{center}
   \includegraphics[scale=.35]{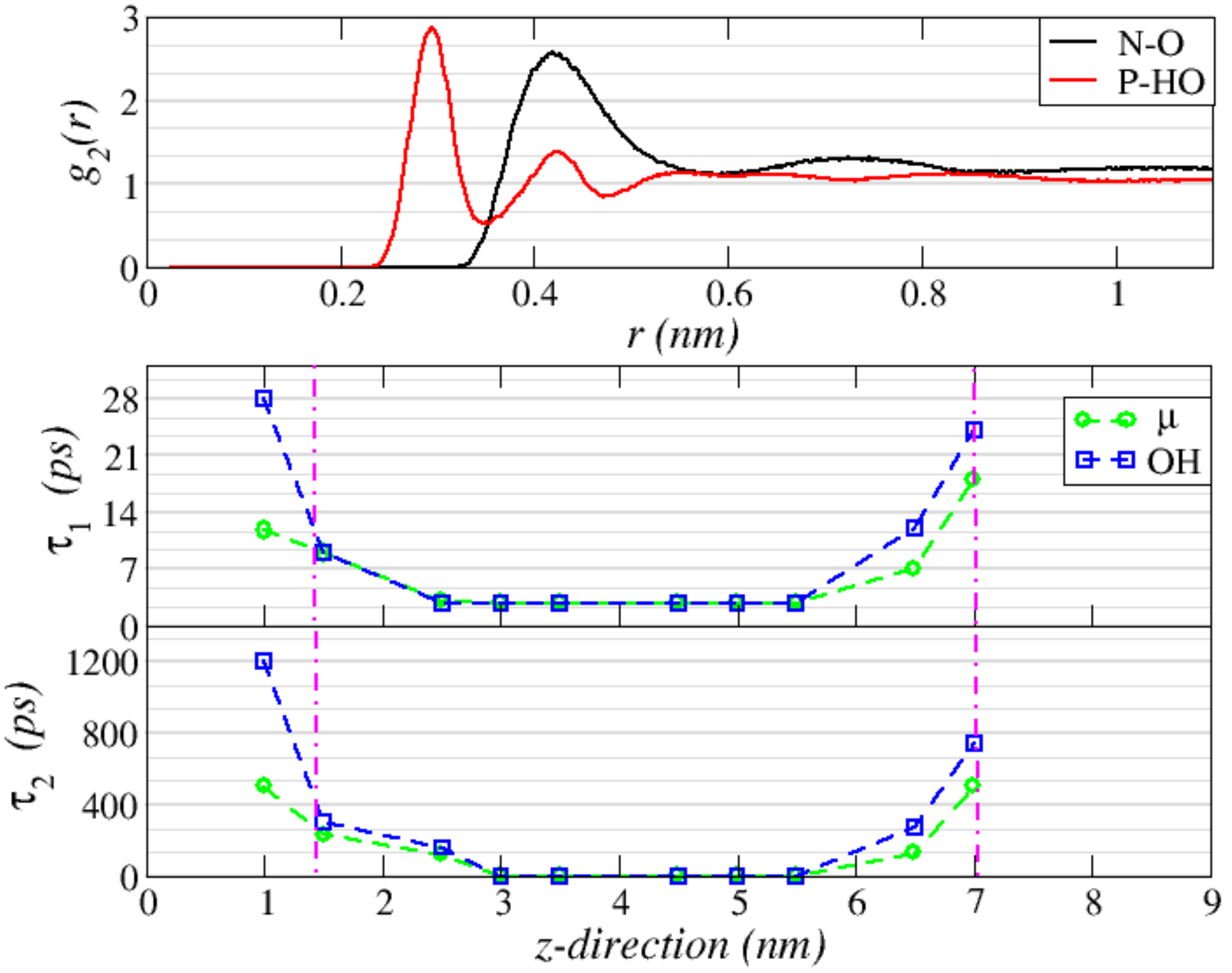}
    \caption{\label{fig:fig2} 
    Upper panel: Two body pair correlation function computed for the N-O and the P-HO vectors in black and red, respectively. Middle and lower panels: \emph{Slow} and \emph{very slow} relaxation times  $\tau_1$ and $\tau_2$, respectively,  computed for the $\mu$ (green open circles) and for the OH (blue open squares) vectors, as a function of the distance from the surface. The magenta lines define the average position of the water-lipid fluctuating surfaces.}
  \end{center}
\end{figure}

In order to further rationalize the interactions between hydration water and phospholipid heads, we computed~\cite{martelli_fop} the correlation function
\begin{equation}
    C_{\bm{\delta}}(t)\equiv\left< \bm{\delta}(t)\cdot\bm{\delta}(0) \right>
    \label{Cdelta}
\end{equation}
where $\bm{\delta}$ is the N-O vector or the P-HO vector. Interestingly, we have found that the P-HO vector has a longer lifetime compared to the N-O vector, indicating that the interactions between P and water hydrogen atoms are stronger than the interactions between N and O~\cite{martelli_fop}. This conclusion is consistent with the observation that the P-HO two body pair correlation function is characterized by a first peak at  a distance shorter than the N-O two body pair correlation function (Fig.~\ref{fig:fig2} upper panel).

Starting from the observation that the N-O and the P-HO vectors have different lifetimes, we  hypothesized that such difference can have an effect on the rotational dynamics of hydration water. In particular, we  supposed that the rotations around the water dipole moment $\bf{\mu}$ are different with respect to the rotations around $\overrightarrow{\rm OH}$ vector. In Ref.~\cite{martelli_fop}, we  computed $C_{\hat{\mu}}$ and $C_{\overrightarrow{\rm OH}}$ and we fit the two correlation functions with a double exponential, with characteristic times  $\tau_1$ and $\tau_2$, that intuitively reveals the effects of the electrostatic interactions on the slow relaxation. We  calculated the relaxation times $\tau_1$ and $\tau_2$ in bins parallel to the membrane surface and centered at increasing distances from the membrane (Fig.~\ref{fig:fig2}, middle and lower panels). 

We found that the \emph{slow} relaxation time, $\tau_1$, is orders of magnitude smaller than the \emph{very slow} relaxation time, $\tau_2$. In particular, approaching the membrane, the $\overrightarrow{\rm OH}$ vector relaxes slower than to the $\hat{\mu}$ vector. 
This is in agreement with the finding that the P-HO interaction is stronger than the N-O interaction. This result can be rationalized by observing that the lipids have different (delocalized) charges on the N-heads and on the P-functional groups and that these charges affect the rotation of water around the two vectors in different way. 

The slowing down of the rotational degrees of freedom (Fig.~\ref{fig:fig2}) decreases upon increasing the distance from the membrane surface. In particular, at distances of $\sim1.3$~nm from the membrane the relaxation times for the $\hat{\mu}$ vector and for the $\overrightarrow{\rm OH}$ vector become indistinguishable, as expected in bulk water. 

In view of the very high values of the relaxation times in the proximity of the membrane, we  hypothesized that the electrostatic interactions with phospholipid heads might cause a slow down in the diffusivity of water molecules  comparable --and hence measurable-- with that of water at low temperatures~\cite{martelli_fop}. 
%We showed~\cite{martelli_fop}  that the overall diffusion of water in the proximity of the surfaces is, indeed, comparable to the diffusion in water at supercooled conditions. 
%In order to draw such conclusion, 
To check our hypothesis,
we measured the standard displacement of water molecules in terms of bond units (BU), defined as the distance traveled by water molecules normalized with respect to the oxygen-oxygen mean distance (which is a temperature-independent quantity), and we  compared it with the same quantity for water at supercooled conditions. For a large enough simulated time, a standard displacement of $<1$ BU would correspond to water molecules rattling in the cage formed by their nearest neighbors. This case would represent a liquid in which the translational degrees of freedom are frozen.

We found that, in the proximity of the membrane surface, water molecules suffer from a dramatic slow down of $\sim60\%$ with respect to the value of bulk water at biological thermodynamic conditions. 
Moreover, upon increasing the distance from the lipid heads, we found that bulk diffusivity is recovered at $\sim1$~nm, the domain of definition of hydration water. 
Considering that the diffusivity of water close to the lipid heads is comparable with that of water at supercooled conditions, we concluded that such a slow-down could be interpreted effectively as a reduction of the thermal energy of water~\cite{martelli_fop}.

\section{Structure}
As presented above, the dynamics of bulk water is recovered approximately at $\sim 1.3$~nm away from a membrane. However, as we will discuss in the following, the structure analysis of hydration water~\cite{martelli_fop} shows how long-range interactions spread at  much larger distances,  opening a completely new scenario for the understanding of water-membrane coupling. In particular, we analyzed~\cite{martelli_fop} how the water intermediate range order (IRO) changes moving away from  a membrane.

Modifications in the connectivity of disordered materials induce effects that extend beyond the short range. This is, for example, the case for amorphous silicon and amorphous germanium~\cite{www}.   
Likewise, at specific thermodynamic conditions, water acquires structural properties that go beyond the tetrahedral short range and are comparable to that of amorphous silicon~\cite{martelli_hyperuniformity}.

In Ref.~\cite{martelli_fop}  we  adopted a sensitive local order metric (LOM) introduced by Martelli et al.~\cite{martelli_LOM} to characterize local order in condensed phase. The LOM provides a measure of how much a local neighborhood of a particle $j$ ($j=1, \dots , N$) is far from the ground state. For each particle $j$, the LOM
maximizes the spatial overlap between the $j$ local neighborhood, made of $M$ neighbours $i$ with coordinates $\mathbf{P}_i^j$ ($i=1, \dots , M$),  and a reference structure --the ground state-- with coordinates $\mathbf{R}^j$. 
The LOM is defined as:
\begin{equation}
    S(j)\equiv \max_{\theta,\phi,\psi;\mathcal{P}}\prod_{i=1}^{M}\exp\left(-\frac{\left|\mathbf{P}_{i_\mathcal{P}}^j-\mathbf{R}^j\right|^2}{2\sigma^2M}\right)
    \label{lom}
\end{equation}
where  $(\theta, \phi, \psi)$ are the Euler angles for a given orientation of the reference structure $\mathbf{R}^j$,  $i_\mathcal{P}$ are the indices of the neighbours $i$
 under the  permutation $\mathcal{P}$,   $\sigma$ is a parameter
representing the spread of the Gaussian domain. 
The parameter  $\sigma$  is chosen such that the tails of the Gaussian functions stretch to half of the O-O distance  in the second coordination shell of $j$ in the  structure $\mathbf{R}^j$. As reference $\mathbf{R}^j$, we choose the ground state for water at ambient pressure, i.e. cubic ice. 
The site-average of Eq.~(\ref{lom}), 
\begin{equation}
    S_C\equiv\frac{1}{N}\sum_{j=1}^N S(j),
    \label{score}
\end{equation}
is by definition the \emph{score function} and 
gives a global measure of the symmetry in the system with respect to the reference structure.
The LOM and the score function has provided physical insights into a variety of systems~\cite{martelli_searching,martelli_unravelling_2019,santra_bnnt}, hence they are particularly suitable also 
to characterize~\cite{martelli_fop} and quantify~\cite{martelli_acsnano} how far
the  membrane affects the water structural properties.

\begin{figure}%[ht]
  \begin{center}
   \includegraphics[scale=.35]{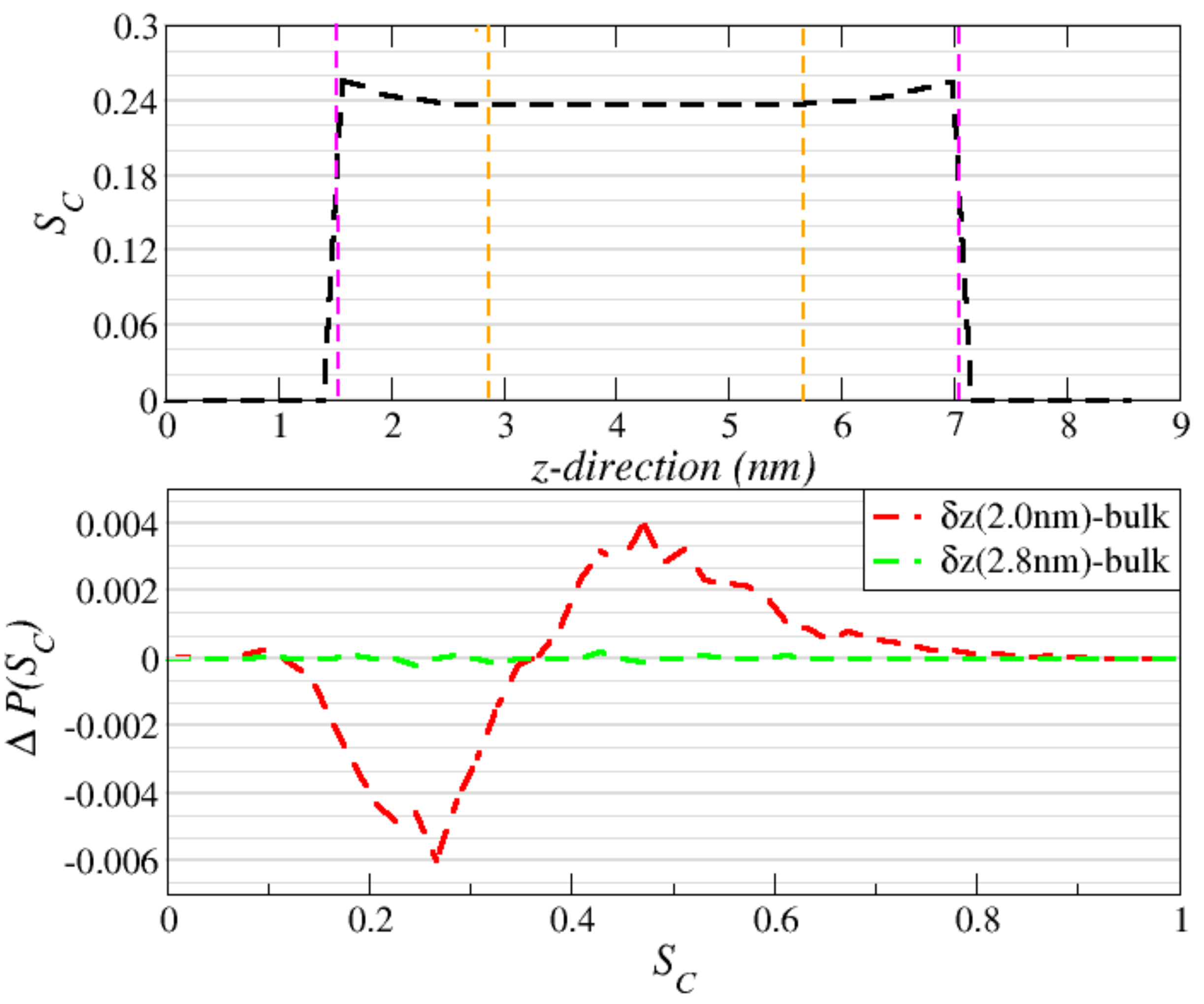}
    \caption{\label{fig:fig3} 
    Score function $S_C$ for water between DMPC membrane leaflets. Vertical magenta lines indicate  the average positions of the water-lipid interfaces. The majority of water is, on average, in the range between $z=1.5$ and 7 nm.
    Upper panel: $S_C$ of water molecules belonging
    to a bin centered at distance $z$ from the center of the lipid bilayer at 0 and with a bin-width of 1/10 of the entire system. 
    Vertical dashed orange lines mark the region where $S_C$ approaches the value in bulk water. 
    Lower panel: Water reaches the  $S_C$ bulk value only at $\approx 2.8$~nm away from the water-lipid interfaces, as shown by the 
    difference $\Delta P(S_C)$ between the probability density distribution $P(S_C)$ for bulk water and that at a specific distance $\delta$ from the membrane. Here we show $\Delta P(S_C)$ 
    for  $\delta z=2.0$~nm (red line), with the bin centered at $z=3.5$~nm,  and for  $\delta z=2.8$~nm (green line), with the bin centered at $z=4.3$~nm.}
  \end{center}
\end{figure}

We found~\cite{martelli_fop}  that the overall score function, Eq.~(\ref{score}), for water tends to increase at very short distances from the membrane and is comparable to bulk at $\gtrsim 1.3$ nm away from the membrane 
(Fig.~\ref{fig:fig3} upper panel). 
The IRO enhancement is not dramatic, but can not be simply discarded. 

Hence, both the dynamics and the IRO are affected as far as $\approx 1.3$ nm away from the membrane.
Therefore, in Ref.~\cite{martelli_fop} we  proposed that the dynamical slow-down and the enhancement of the IRO are two effects related to each other. 
We  suggested that the dynamical slow-down corresponds to an effective reduction of thermal noise that, ultimately, allows water molecules to adjust in slightly more ordered spatial configurations in the proximity of the membrane. 

Moving away from the membrane, at distances $\gtrsim1.3$~nm, $S_C$ seems to reach a plateau, suggesting that a convergence to the bulk value should fall into the distance domain of hydration water. 
To check this, we computed the probability density distribution $P(S_C)$ of Eq.~(\ref{score}) in the bin 
%the farthest from both lipid surfaces, and 
centered at $\delta z= 2$~nm away from the surfaces ($z=3.5$~nm), and we compared it with the distribution of $S_C$ computed in a box of bulk water at the same thermodynamic conditions (Fig.~\ref{fig:fig3} lower panel). 

Surprisingly, the two distributions \emph{do not} overlap. This result indicates that the membrane perturbs the structure of water at the intermediate range of, at least, $\sim1.6$~nm, considering half bin-width. This distance is much larger than that defining hydration water. 

We found~\cite{martelli_acsnano} an overlap between the bulk-water distribution  and that for the confined water  only if between the two membrane leaflets there is enough water to reach distances as far as $\delta z= 2.8$~nm from the membrane. 
Such a remarkable result indicates that the membrane affects the structural properties of water at least as far as $\sim2.4$~nm, accounting for the $\sim0.4$~nm half bin-width.
This distance can be considered twice the domain of definition of hydration water. 

Therefore, the definition of hydration water, as well as its role, should be extended to account for the repercussion of the membrane on the water structure. Or it should be revised, in order to further re-define its concept. In order to properly frame our observations into a consistent picture, in addition to our structural analysis of the membrane effects on the water-O positions, we have analyzed next the topology of the hydrogen bond network (HBN) which provides another measure of the IRO, but from the perspective of the HBs.

\section{Network topology}

The properties of network-forming materials are governed by the underlying network of bonds~\cite{martelli_rings}. However, the topology of this network is very rarely investigated because of the difficulty of such analysis. 

A possible approach is through the \emph{ring statistics}. It consists in defining, characterizing and counting the number of closed loops that  are made of links (or bonds) between the vertices of the network. The ring statistics allows to study, in particular, the network topology of amorphous systems~\cite{leroux_ring,yuan_efficient}, clathrate hydrates~\cite{chihaia_molecular}, and chalgogenide glasses~\cite{blaineau_vibrational}. It is, also, an essential tool to characterize continuous random networks~\cite{www,wooten_structure,djordjevic_computer,barkema_event,barkema_high,hudson_systematic}.
 
After some hesitant debut in the field of water~\cite{martonak_2004,martonak_2005}, ring statistics has been embraced more and more as a tool to study water properties, starting from its application by Martelli et al. to characterize the transformations in the bulk water HBN near the liquid-liquid critical point~\cite{martelli_nature}. 
Since then, ring statistics has been an essential tool for investigating the properties of  water in its liquid  phase~\cite{santra_2015,martelli_rings,camisasca_proposal}, as well as its amorphous states~\cite{martelli_searching,martelli_rings,martelli_LOM}, and for inspecting the dynamics of homogeneous nucleation~\cite{russo_2014,leoni_2019,fitzner_ice}. 

Based on the idea that the connectivity in network-forming materials governs theirs properties, we explored how the topology of the HBN changes when water is confined between phospholipid membranes~\cite{martelli_acsnano}. 
In fact, the HBN is what differentiates water from "simple" liquids~\cite{pauling}. 

In water the HBN is directional. Hence, there are several ways for defining and counting rings. Martelli et al. showed that each of these possibilities carries different, but complementary, physical meaning~\cite{martelli_rings}. 

Here we use a definition for the HB that was initially introduced by Luzar and Chandler~\cite{chandler_HB} and is common in the field.
However, other definitions are possible, due to our limited understanding of the HBs. Nevertheless, it has been shown that all these definitions have a satisfactory qualitative agreement  over a wide range of thermodynamic conditions~\cite{prada_2013,shi_2018_2}. 

\begin{figure}%[ht]
  \begin{center}
   \includegraphics[scale=.50]{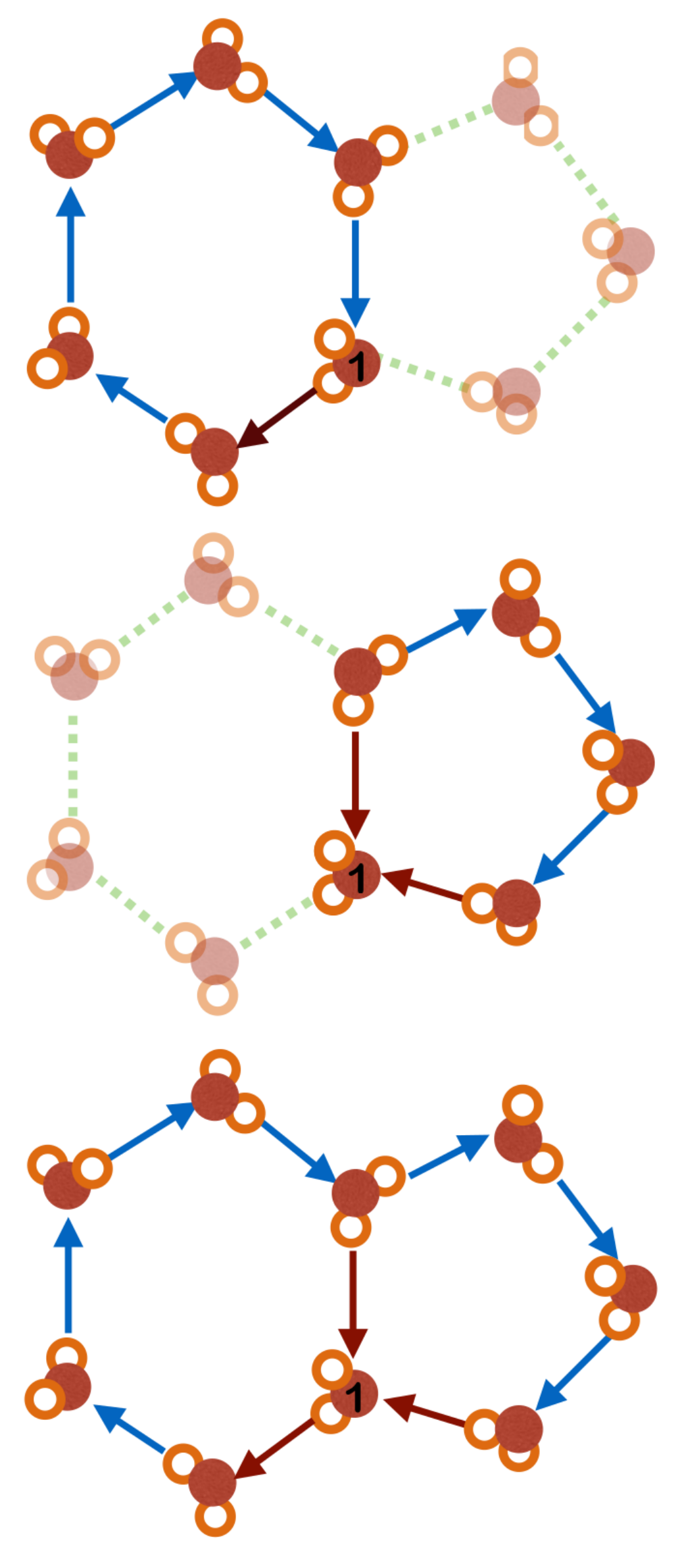}
    \caption{\label{fig:fig4} Schematic representation of 
three possible ways of defining the rings in the water directional network.
In each case, we start counting from the 
water molecules labeled as 1, with O atoms in solid brown and H atoms in white, and 
we follow the directional HBs from H to O (arrows) along the HBN, until we return to molecule 1 or until we exceeds 12 steps. 
We consider only rings that cannot be decomposed into sub-rings. 
Top: A ring is formed only when molecule 1 donates HBs (brown arrow). In the example, the shortest ring is the hexagonal one (blue arrows). 
Center: A ring is formed when molecule 1 donates or accepts (brown arrows) HBs. In the example, the shortest ring is the  pentagonal ring (arrows). 
Bottom: Any ring formed by molecule 1 is considered, starting from any of its HBs (brown arrows), without bond or ring's length constraints. In the example, there are a hexagonal and a pentagonal ring.  Martelli et al. adopted the latter definition in Ref.~\cite{martelli_acsnano}.
    }
  \end{center}
\end{figure}

In Fig.~\ref{fig:fig4} we present three possible ways of defining rings in a directional network, as in the case of water. The first (Fig.~\ref{fig:fig4} Top)  explicitly looks for the shortest ring~\cite{king} starting from the molecule 1, when this molecule donates one HB, regardless whether other molecules  in the ring accept or donate a bond. This definition emphasizes the intrinsic directional nature of the HBN. The second definition (Fig.~\ref{fig:fig4} Center) considers only the shortest ring formed when molecule 1 can only accept a HB. The third definition (Fig.~\ref{fig:fig4} Bottom), adopted by Martelli et al.~\cite{martelli_rings}, ignores both the donor/acceptor nature of the starting molecule and the shortest-rings restriction, leading to a higher number of rings. The reader can refer to the original work~\cite{martelli_rings} for further details about the definitions  and their physical meaning in the case of bulk liquid and glassy water at several thermodynamic conditions.

\begin{figure}%[ht]
  \begin{center}
   \includegraphics[scale=.27]{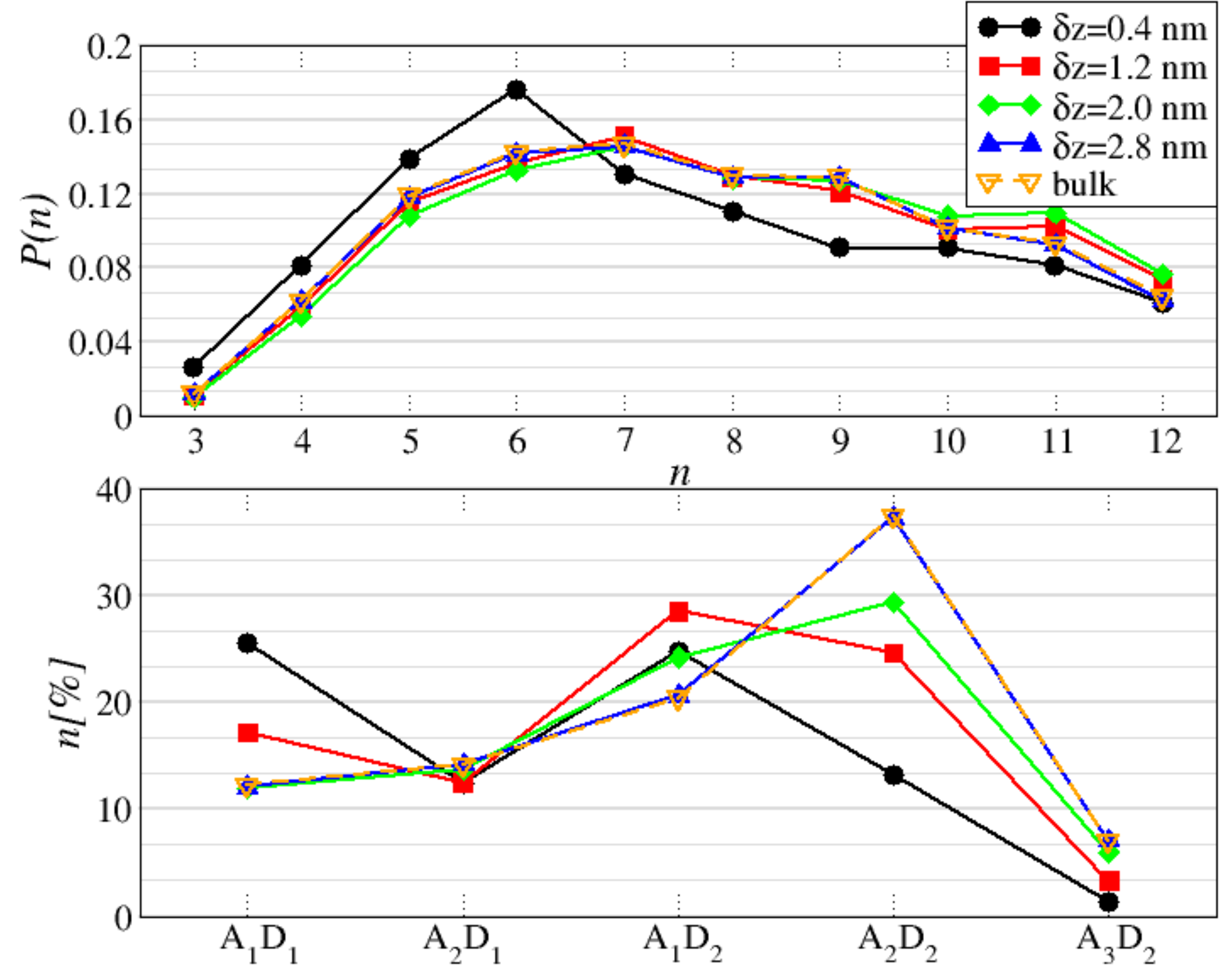}
    \caption{\label{fig:fig5} HBN ring statistics at a distance $z$ from the average position of the fluctuating membrane and in bulk water. In both panels the sets of data are for bulk water (open orange triangles), and 
 $z=0.4$~nm (black dots), $1.2$~nm (red squares), $2.0$~nm (green diamonds), and $2.8$~nm (blue triangles).  Quantities at a given distance from the membrane are calculated in $0.8$~nm-wide bins centered at $z$.  
    Upper panel: Probability of having $n$-member rings, $P(n)$.    
    All $P(n)$ are normalized to unity and, therefore, do not reflect the total number of rings of a given size. 
    Lower panel: Percentage-wise decomposition of the HBs per water molecule into acceptor-(A) and donor-(D).  
    The $x$-axis labels $\textit{A}_x\textit{D}_y$ indicate the number of acceptor ($\textit{A}_x$) and donor ($\textit{D}_y$) HBs, respectively, of the configurations schematically represented in the plot (with the oxygen of central water molecule in blue). 
    For clarity we omit combinations  with minor contributions, \textit{e.g.}, $\textit{A}_3\textit{D}_1$, $\textit{A}_0\textit{D}_y$, and $\textit{A}_x\textit{D}_0$.
    }
  \end{center}
\end{figure}
The  authors of Ref.~\cite{martelli_acsnano}  computed  the probability of having a $n$-folded ring, $P(n)$, as a function of the distance $z$ from the membrane.  They found that near  the membrane the $P(n)$ is strikingly different from that of bulk water (Fig.~\ref{fig:fig5}, upper panel). 
In particular, the distribution is richer in hexagonal and shorter rings and is poorer in longer rings. 

This result points towards two main conclusions: 
(i) 
For membrane-hydration water, at a distance $z\leq 0.8$~nm, the HBN tends to  be preferentially  ice-like, i.e.,  dominated by hexagonal rings. 
This observation  is consistent with the results, discussed in the previous sections, showing  
that membrane-vicinal water is characterized by enhanced IRO and slower dynamics than bulk water. 
(ii) 
The reduced number of longer rings in the hydration water is consistent with the reduction of the overall dimensionality of the system due to the interface. The membrane fluctuating surface  reduces the available space for the HBN in the first layer of hydration water. 

All the $P(n)$ calculated at larger distances, $z>0.8$~nm, are quite different from that for the hydration water and  gradually converge towards a the bulk case upon increasing $z$. In particular, the probability of hexagonal rings decreases progressively,  while longer rings become more and more frequent. 

This sudden change in $P(n)$, between the first and the following bins, is consistent with the results, discussed in the previous sections, demonstrating the existence of a drastic change in structure and dynamics between bound water, in the first hydration layer, and unbound water, away from the membrane~\cite{calero_membranes_2019}.  
Here, the border between the two regions is increased from $\sim 0.5$~nm~\cite{calero_membranes_2019} to 
$\sim 0.8$~nm due to the membrane fluctuations, that are not filtered out in  Ref.~\cite{martelli_acsnano}, and to the spatial resolution, i.e., the bin-size, of the analysis.

The HBN of bulk water is finally recovered in the bin centered at $z=2.8$~nm away from the membrane,
i.e., for $z\geq 2.4$~nm. Remarkably, this distance corresponds to the same  at which  water recovers the IRO of bulk water~\cite{martelli_fop}, as discussed in the previous section. This important result indicates a clear connection between the structural properties of water molecules and the topology of the HBN, while further pointing toward the necessity of revising the concept of hydration water.

The quality of the HBN, in terms of broken and intact HBs, is a tool of fundamental importance to fully cast the topology of the HBN in a consistent and complete physical framework. 
As a matter of fact, the presence of coordination defects affects the fluidity of water and is directly related to its capability of absorb long range density fluctuations~\cite{martelli_hyperuniformity}. 
Therefore, the authors in Ref.~\cite{martelli_acsnano}  complemented their investigation of the HBN topology with the analysis of its quality. 

They decomposed the HBs per water molecule into acceptor-(A) and donor-(D) types (Fig.~\ref{fig:fig5} lower panel). 
They label as $\textit{A}_2\textit{D}_2$ a water molecule with perfect coordination, i.e., donating two bonds and accepting two bonds and as $\textit{A}_x\textit{D}_y$ the others  accepting $x$   and donating $y$ bonds.
%In the lower panel of fig.~\ref{fig:fig5}, the computed  ratio of water molecules that have different coordination, i.e., are not in the $\textit{A}_2\textit{D}_2$ configuration at  distance $z$ from the membrane. 
%
They focused their attention on the following coordination configurations: 
$\textit{A}_1\textit{D}_1$, $\textit{A}_2\textit{D}_1$, $\textit{A}_1\textit{D}_2$, $\textit{A}_2\textit{D}_2$ and $\textit{A}_3\textit{D}_2$, as other configurations do not contribute significantly. 

First, they checked that in bulk water, at ambient conditions, the predominant configuration is  $\textit{A}_2\textit{D}_2$. For the TIP3P model of water, this configuration accounts for $\sim35\%$ of the total composition.
The second most dominant configuration in bulk is  
$\textit{A}_1\textit{D}_2$ with $\sim20\%$, followed by 
$\textit{A}_2\textit{D}_1$ with $\sim13\%$,  
$\textit{A}_1\textit{D}_1$  with $\sim12\%$ and, finally,  
$\textit{A}_3\textit{D}_2$  accounting for less then $10\%$ (Fig.~\ref{fig:fig5} lower panel).

Such distribution qualitatively reflects the distribution in \emph{ab initio} liquid water at the same thermodynamic conditions~\cite{distasio_2014}. 
Hence, it suggests that classical potentials can carry proper physical information even in very complex systems such as biological interfaces. 

In the proximity of the membrane, the network of HBs largely deviates from that of bulk water, except for  the under-coordinated configuration $A_\textit{2}D_\textit{1}$. 
In particular, the coordination defects
$A_\textit{1}D_\textit{1}$ and 
$A_\textit{1}D_\textit{2}$ dominate the distribution, with $\sim25\%$ each, followed by the configurations
$A_\textit{2}D_\textit{1}$ and 
$A_\textit{2}D_\textit{2}$, with  $\sim15\%$ each, and a minor percentage of higher coordination defects $A_\textit{3}D_\textit{2}$, with $\sim3\%$. 

However, the small percentage of perfectly coordinated configurations, $A_\textit{2}D_\textit{2}$, near the membrane seems inconsistent with the higher local order observed at the same distance~\cite{martelli_fop,martelli_acsnano}, and with  the enhanced hexagonal ring-statistics of the HBN~\cite{martelli_acsnano}, already discussed.
Such discrepancy is only apparent for the following two reasons. 

First, both the structural score function, $S_C$, and the ring statistics are a measure of the IRO beyond the short range. On the contrary, the quality of the HBN, in terms of defects, is  a measure only of the short range order.

Second, the defects analysis includes only HBs between water molecules and do not account for the strong HBs between water molecules and the phospholipid headgroups. Instead, as  discussed in the previous section~\cite{calero_membranes_2019}, $\sim30\%$ of the water molecules in the first hydration shell are bound to the membrane with at least one HB.

Away from the membrane, upon increasing the distance, Martelli et al.~\cite{martelli_acsnano}  observed a progressive enhancement of perfectly tetra-coordinated configurations (Fig.~\ref{fig:fig5} lower panel).
They found a progressive depletion of all coordination defects, up to recovering the bulk-water case at distance $z\geq 2.4$~nm from the membrane, as for the probability distribution of $S_C$ and the HBN topology.

The intriguing evidence that the under-coordinated defect $A_\textit{2}D_\textit{1}$ remains almost constant at all distances is, for the moment, not explained. Indeed, it could be due to a variety of reasons, going from the presence of water-membrane HBs in the first hydration layer, to the propagation of defects in bulk, and it would require a detailed study.

\section{Conclusions and future research directions}

The results summarized in this short review question our common understanding of hydration water near soft membranes, such as those in biological systems. This water layer, often called bio-water, is usually  considered as $\sim1$~nm wide 
and is regarded as the amount of water that directly shape and define the biological activity in proteins, cells, DNA, etc. 
Such definition has been proposed based on results, both from experiments and computations, showing that the water dynamics and density are affected by the  biological interface within $\sim1$~nm, while they recover the bulk behavior  at larger distances. 

In our calculations based on well-established models of water nanoconfined between DMPC membranes, instead, we found new evidences that indicate the need for a revised definition of hydration water. We achieved this conclusion by focusing on physical quantities that have been not thoroughly, or not at all,  considered before.

In particular, by considering the instantaneous local distance of water from the membrane, 
Calero and Franzese were able to unveil the existence of a new interface between bound and unbound water $\sim 0.5$~nm away from the membrane-water interface~\cite{calero_membranes_2019}. 
Bound water behaves like a structural component of the membrane and has a translational and rotational dynamics that is intermediate between water inside and outside the  membrane~\cite{calero_membranes_2019}.  
Bound-water dynamics is dominated by the strong HB with the membrane and is orders of magnitude slower than the unbound water. The dynamics of bulk water is recovered only $\sim 1.3$~nm away from the membrane.

However, we showed that the membrane interface has an effect on the structure of the hydration water at a distance almost twice as large, up to, at least, $\sim 2.4$~nm~\cite{martelli_fop}. We got such a result by analyzing how the water structure, and its IRO, changes by moving away from the membrane. To this goal, we evaluated the score function, a structural observable that quantifies how close is the local structure to a reference configuration, in our case the cubic ice.  Also in this case, we found that water $\sim 1.3$~nm away from the membrane has a small but measurable IRO enhancement. Hence, within this range both the dynamics and the structure of hydration water undergo an effective reduction of the thermal noise, that we interpret  as a consequence of the interaction with the membrane.
Also, we have shown that different chemical species constituting the lipid heads interact with water molecules with different strengths, hence providing a rationale for the contributions to the observed dynamical slow-down in the proximity of the surface~\cite{martelli_fop}.

Furthermore, Martelli et al.~\cite{martelli_acsnano} analyzed the IRO from the HB perspective by studying the HBN topology and its ring statistics. They found that water within $\sim 0.8$~nm from the average position of the fluctuating membrane has an excess of hexagonal and shorter rings, and a lack of longer rings, with respect bulk water. Moreover, the defect analysis of the HBN showed that water in this  $\sim 0.8$~nm-wide layer has a lack of  water-tetra-coordinated  molecules and an excess of water bi-coordinated molecules. This result does not contradict the enhanced water IRO within the same layer,  because the HBN defects analysis measures only the short range order and does not account for the water-membrane HBs.

Martelli et al.~\cite{martelli_acsnano} found also a sudden change in the HBN around  $0.8$~nm, with a ring statistics that approaches that of bulk. This result confirms the qualitative difference between bound and unbound water~\cite{calero_membranes_2019}. 

The analysis of the HBN ring statistics and the HBN defects
 show that the membrane interface generates a perturbations in the ring statistics that extends as far as, at least, $\sim 2.4$~nm~\cite{martelli_acsnano}. These observations, therefore, corroborate that  the water structure is affected by the membrane interface up to a distance at least  twice as large as that usually associated to the hydration water. 

%Interestingly, the investigations of the HBN  shows that the percentage of coordination defects of the kind $A_\textit{2}D_\textit{1}$  at increasing distances from the surface, is independent of the distance from the membrane. Such observation remains, to date, still unexplained, but we speculate here that could be related to long range perturbations in the dipole of water molecules.

All these findings should be taken into account when interpreting experimental results and when developing membrane-water interaction potentials. 
They can help in better understanding water in biological processes at large and, in particular, those phenomena where hydration plays a role. From a more general perspective, these calculations imply that the concept of hydration should be revised in order to account for the  results presented here.

Our conclusions entail further investigation about the relationship between 
diseases, possibly promoted by extracellular matrix variations, e.g., of hydration or ionic concentration, with the water HBN rearrangements. Example of such illness are cardiac disease and arterial hardening in healthy men~\cite{Arnaoutis:2017aa}, or atherosclerosis and inflammatory signaling in endothelial  cells~\cite{10.1371/journal.pone.0128870}.  Indeed, variations of ionic concentration drastically change the water HBN structure~\cite{Mancinelli:2007fk} and  dynamics~\cite{Fayer:2009zx}, with an effect that is similar to an increase of pressure~\cite{Gallo:2014ab}. While dehydration has consequences on the dynamics and the structure of the water near a membrane that resemble those of a temperature decrease~\cite{calero_membranes_2019}. 

In particular, we foresee the extension of these calculations to out-of-equilibrium cases.
 %Nonetheless, so far molecular simulations of membranes involved equilibrium states. 
%On the other hand,
Indeed,  it has been recently shown that the potency of antimicrobial peptides may not be a purely intrinsic chemical property and, instead, depends on the mechanical state of the target membrane~\cite{losasso_2019}, which varies at normal physiological conditions.

%The use of the instantaneous local distance $\xi$ gives the possibility to explore the interesting comparison between soft and hard confinement. 
%On the other hand, translational and rotational dynamics of water are affected also when water is under hard confinement. Such results are in agreement and extend previous simulation works studying water orientational relaxation as a function of the hydration level of lipids~\cite{Zhang_Berkowitz_JPhysChemB2009}. 
%Along this line, Calero and Franzese have recently shown that the water translational and rotational dynamics exhibit an oscillatory behaviour when confined in an open graphene slit-pore down to the nano-meter scales, due to the layering of water~\cite{calero_2020}. 

%Our results provide a rationale for understanding how different components of the lipid heads interact with water molecules and contribute to the dynamical slow down. Such detailed reasoning could be applied to other systems such as membrane of biological interest and/or with mixed composition. 

\begin{acknowledgments}
F.M. acknowledges support from the STFC Hartree Centre's Innovation Return on Research programme, funded by the Department for Business, Energy and Industrial Strategy. C.C. and G.F. acknowledge the support of Spanish grant PGC2018-099277-B-C22 (MCIU/AEI/ERDF), and G.F. the support by ICREA Foundation (ICREA Academia prize).
\end{acknowledgments} 

%\nocite{*}
%\linespread{0.1}
%\bibliographystyle{unsrt}
\bibliography{main}% Produces the bibliography via BibTeX.

\end{document}